\documentstyle[amstex,amsfonts,aps,prb,overcite,draft]{revtex}

\begin{document}
\title{Multiresolution analysis in statistical mechanics. I. Using wavelets to
calculate thermodynamic properties}
\author{Ahmed E. Ismail, Gregory C. Rutledge, and George Stephanopoulos}
\address{Department of Chemical Engineering, Massachusetts Institute of Technology,\\
Cambridge, MA\ 02139}
\date{\today}
\maketitle

\begin{abstract}
The wavelet transform, a family of orthonormal bases, is introduced as a
technique for performing multiresolution analysis in statistical mechanics.
The wavelet transform is a hierarchical technique designed to separate data
sets into sets representing local averages and local differences. Although
one-to-one transformations of data sets are possible, the advantage of the
wavelet transform is as an approximation scheme for the efficient
calculation of thermodynamic and ensemble properties. Even under the most
drastic of approximations, the resulting errors in the values obtained for
average absolute magnetization, free energy, and heat capacity are on the
order of 10\%, with a corresponding computational efficiency gain of two
orders of magnitude for a system such as a $4\times 4$ Ising lattice. In
addition, the errors in the results tend toward zero in the neighborhood of
fixed points, as determined by renormalization group theory.
\end{abstract}

\section{Introduction}

Spin models are popular tools for theoretical calculations and for numerical
simulations, as their universality classes allow a huge range of different
systems---as varied as binary metal alloys, surface adsorption, and neural
networks---to be modeled simultaneously. For example, even the ``trivial''
one-dimensional Ising model can be used to model the helix-coil transition
in biopolymers; the deep connection between magnetic models and polymer
chains allow us to predict scaling behavior and other properties across an
even wider range of materials.\cite{Freed87} Lattice models are still widely
used in modeling the thermodynamics of complex systems, because their
regular structure simplifies the type and nature of interactions among
components of the system. Moreover, the difficulty in obtaining analytical
solutions of lattice systems, and the relative ease of computational
simulations thereof, make them ideal test cases for new simulation
algorithms.

Although simulations of lattice models are relatively straightforward to
implement, they share the same drawbacks as off-lattice models. The chief
drawback is that as the number of particles grows large, the time required
to sample the system accurately increases rapidly. A popular approach for
addressing this problem is to {\em coarse-grain} the system: that is, we
``rescale'' the problem by increasing the basic size of a simulation
element. For example, we might coarse-grain an atomic representation of a
polymer chain into a ``united atom'' model, where a chain molecule is
treated as if it consisted only of the backbone. More creative approaches
redefine the problem to be addressed:\ for example, Mattice and coworkers
have produced a method which maps a polymer chain atomistically onto a
high-coordination lattice; this lattice is then used as the basis for a
Monte Carlo simulation; the resulting configuration is then used to map back
to continuous space to provide ``evolution in time.'' \cite
{Rapold96,Cho97,Doruker97,Doruker99}

This paper illustrates the use of the wavelet transform as a mathematical
basis for performing thermodynamic computations of lattice models. The
wavelet transform is an important tool in {\em multiresolution analysis},
which analyzes a system simultaneously at several length or frequency scales
selected to reflect the actual physical processes underlying the observed
behavior as closely as possible. The wavelet transform possesses a number of
convenient properties, including orthogonality, compactness, and
reconstruction; we will make these concepts more precise in Section \ref
{ref01} below. The orthogonal nature of most wavelet constructions makes
them a logical choice for use in {\em ab initio} density-matrix quantum
chemistry computations, in which the selection of an accurate basis set is
crucial to the convergence and efficiency of the calculations.\cite
{Arias99,Beylkin99,Johnson99} Wavelet decompositions have been applied
principally in electrical engineering, particularly in the field of signal
processing. In this context, white noise and Markov processes have been
studied using multiscale methods.\cite{Luettgen93,Luettgen93b} To date,
however, wavelet analysis does not seem to have been extensively applied to
models in statistical mechanics. Huang uses wavelet analysis to observe the
statistical distribution of multiplicity fluctuations in a lattice gas,\cite
{Huang97} while Gamero {\em et al.} employ wavelets to introduce their
notion of {\em multiresolution entropy}, although their primary goal is
dynamic signal analysis rather than statistical mechanics simulations,\cite
{Gamero97} while O'Carroll attempts to establish a theoretical foundation
connecting wavelets to the block renormalization group.\cite
{Ocarroll93,Ocarroll93b} A more in-depth review of the connection between
wavelets and renormalization theory is provided in a recent monograph by
Battle.\cite{Battle99}

\section{Wavelet transform fundamentals}

\subsection{The conceptual picture}

The wavelet transform is a hierarchical method for decomposing a data set
into averages and differences. Like the Fourier transform, it can be used to
provide a decomposition in both real space and reciprocal space (${\bf k}$%
-space), or time space and frequency space. Unlike the Fourier transform,
however, it is capable of providing simultaneously localized transformations
in both real and reciprocal space. A function localized in position space,
such as a finite impulse function, cannot be represented by a few terms of
its Fourier series: many terms are required before good convergence is
achieved. By contrast, in wavelet space, this same function can be almost
completely described by just a handful of wavelet coefficients. Although the
first wavelet was discovered almost a century ago by Haar,\cite{Haar10} they
have become an important computational technique only in the last decade,
following the work of Mallat,\cite{Mallat89,Mallat89b} Daubechies,\cite
{Daubechies92} and others. \cite{Cohen93,Sweldens94,Sweldens94b}

The wavelet transform, like any other transform, takes a mathematical object
and transforms it into another: we can represent its action by writing 
\begin{equation}
\tilde{u}={\cal W}\left[ u\right] ;  \label{def}
\end{equation}
the specific form of ${\cal W}$ depends both on the type of wavelet we have
selected, and the object $u$ which we wish to transform. All versions of the
wavelet transform ${\cal W}$, however, are derived from the same source: a
set of coefficients which define the transform. If $u$ is a discrete data
set, such as a signal sampled at regular intervals, then ${\cal W}$ is
usually represented as a matrix; while if $u$ represents a continuous data
set, such as the same signal measured at all times, then ${\cal W}$ acts as
an integral operator. While the matrix form of ${\cal W}$ is often called a
``filter bank'' and the integral form a ``wavelet transform,'' we will not
distinguish between them in what follows, as the theory developed here for
discrete lattices and filter banks should carry over to continuous systems
and wavelet transforms essentially unchanged.

Similar to the Fourier transform, the wavelet transform decomposes the
object $x$ into two separate components, as two different functions, a
scaling function $\phi $ and a wavelet function $\psi $, both operate on $x$%
. However, the two functions separate its components not into cosines and
sines, but into averages and differences, with a ``wavelength'' equal to the
``window'' over which the scaling and wavelet functions are nonzero. In
another important distinction, the wavelet transform is {\em recursive}, so
that it can be applied in succession to any set of averages which is
produced using that wavelet transform, to produce another level of averages
and another level of details.

We shall now make the above concepts more mathematically precise. Let us
define $u$ to be a discrete set of samples $u=\left( u\left( 1\right)
,u\left( 2\right) ,\ldots ,u\left( n\right) \right) $. Then applying the
scaling and wavelet functions $\phi $ and $\psi $ to $u$ create a set of
averages $s\left( i\right) $ and a set of differences $\delta \left(
i\right) $: 
\begin{align}
s\left( i\right) & =\sum_{k=0}^{r-1}\phi \left( k\right) u\left( i+k\right) ,
\label{wt1} \\
\delta \left( i\right) & =\sum_{k=0}^{r-1}\psi \left( k\right) u\left(
i+k\right) ,  \label{wt2}
\end{align}
where $r$ is a finite integer which defines the length scale, often referred
to as the ``size of the support,'' over which $\phi $ and $\psi $ are
nonzero. The index $i$ runs from $1$ to $n;$ generally the data set is
padded with zeros to ensure that all sums in (\ref{wt1}) and (\ref{wt2}) are
well-defined, although periodicity is sometimes used instead. \cite{Strang96}
The coefficients $\phi \left( k\right) $ and $\psi \left( k\right) $ in (\ref
{wt1}) and (\ref{wt2}) are related,\cite{Daubechies92,Strang96} and are
central in controlling the features of the wavelet transform. Note the
wavelet transform is inherently redundant: for every sample $u\left(
i\right) $ in the original set $u$, we now have two values, a local average $%
s\left( i\right) $ and a local difference $\delta \left( i\right) $. Since
the new data are simply linear combinations of the original values, it is
superfluous to retain both sets; at the same time, it is obvious that we
cannot simply discard one set of data and recover all the original
information using only the other data set. Instead, we choose to keep only
the odd-numbered $s\left( i\right) $'s and $\delta \left( i\right) $'s,
eliminating the even-numbered samples; this process is called {\em %
downsampling}.\cite{Strang96} Downsampling removes half of the $s\left(
i\right) $ and half of the $\delta \left( i\right) $, regardless of the
length $r$ of the wavelet. Now we are left with $n$ data:\ $s\left( 1\right)
,s\left( 3\right) ,\ldots ,s\left( n-1\right) $ and $\delta \left( 1\right)
,\delta \left( 3\right) ,\ldots ,\delta \left( n-1\right) $. These $n$ data
points can be stored as the level-one wavelet transform $\tilde{u}$ of $u$,
by assigning $s\left( 1\right) ,s\left( 3\right) ,\ldots ,s\left( n-1\right) 
$ to $\tilde{u}^{\left( 1\right) }\left( 1\right) ,\tilde{u}^{\left(
1\right) }\left( 2\right) ,\ldots ,\tilde{u}^{\left( 1\right) }\left(
n/2\right) $, and the corresponding $\delta \left( i\right) $'s as $\tilde{u}%
^{\left( 1\right) }\left( n/2+1\right) ,\ldots ,\tilde{u}^{\left( 1\right)
}\left( n\right) $. [The superscript $\left( 1\right) $ denotes that the
wavelet transform has been applied once to this data set.] We can either
stop at this level of description, or continue by further decomposing the
averages: then the new object $u^{\left( 1\right) }$ to be transformed is $%
u^{\left( 1\right) }\left( 1\right) =\tilde{u}\left( 1\right) ,\ldots
,u^{\left( 1\right) }\left( n/2\right) =\tilde{u}\left( n/2\right) $, and so
on. Note that although $\tilde{u}^{\left( 1\right) }$ contains the averages $%
s\left( i\right) $'s and the differences $\delta \left( i\right) $'s
obtained in the previous step, successive transforms only apply to averages
obtained in the previous step. This process can be repeated until we have
reduced our set of averages to a single point; no further averaging is
possible. We assume henceforth that the data set $u^{\left( k\right) }$ has
been sufficiently downsampled to retain only the minimum data set required.

\subsection{Properties and examples of wavelet and scaling functions}

\label{ref01}Until this point, we have not introduced any specific wavelet
or scaling functions. Before we do so, we note that the choice of a wavelet
transform to apply to a given system usually hinges on the desired
properties which one wishes to include in the transformed data. Three
principal properties are almost universally required for filter banks and
wavelet families:\cite{Daubechies92,Strang96}

\begin{enumerate}
\item  {\em Perfect reconstruction}: No data is distorted by performing
analysis followed by synthesis, so that the only permissible change is a
delay in recovery of the original sample.

\item  {\em Orthogonality}: Wavelets computed at different length scales or
at different spatial locations are mutually orthogonal; thus fluctuations in
the system are localized at the scales where they are most relevant.

\item  {\em Compact support}: Properly designed wavelets are identically
zero except for a finite interval, which means that exact results can be
obtained using only a finite number of terms.
\end{enumerate}

Other properties, such as orthonormality, symmetry in the functional form of
the wavelet or a certain number of vanishing moments, can be taken into
account when constructing the wavelet transform.\cite{Sweldens96}

The two most commonly encountered selections are the Haar and Daubechies
wavelets, named after their respective discoverers. The Haar pair is the
oldest and simplest set of wavelets:\cite{Haar10}\ the coefficients of the
scaling function are $\phi =\left( \phi \left( 0\right) ,\phi \left(
1\right) \right) =\frac{1}{\sqrt{2}}\left( 1,1\right) $, while the
coefficients of the wavelet function are $\psi =\left( \psi \left( 0\right)
,\psi \left( 1\right) \right) =\frac{1}{\sqrt{2}}\left( -1,1\right) $. No
other wavelet can be described with two points, and therefore no other
wavelet has a support as compact as the Haar wavelet. The scaling function $%
\phi $ simply averages the values stored at neighboring points, while $\psi $
finds the difference between those values; the extra factor of $\sqrt{2}$ is
incorporated to ensure orthonormality between overlapping $\phi \left(
k\right) $ and $\psi \left( k\right) $. A simple example of the action of
the Haar wavelet is shown in Fig. \ref{fig1}.

The Daubechies wavelets are a family of orthonormal functions whose
construction was explicitly designed to have orthogonality as well as
vanishing higher-order moments.\cite{Daubechies88} Daubechies was able to
show that the Haar wavelet is in fact the ``first'' member of the Daubechies
family; that is, the Haar wavelet is the Daubechies wavelet with the
shortest support. The second such member has four terms in its definition:\
the scaling function is defined by $\phi =\left( \phi \left( 0\right) ,\phi
\left( 1\right) ,\phi \left( 2\right) ,\phi \left( 3\right) \right)
\allowbreak =\frac{1}{4\sqrt{2}}\left( 1+\sqrt{3},3+\sqrt{3},3-\sqrt{3},1-%
\sqrt{3}\right) .$ The wavelet function reverses the order of the
coefficients and inverts the sign of every other component, which allows the
orthonormality properties to be satisfied: $\psi =\left( \psi \left(
0\right) ,\psi \left( 1\right) ,\psi \left( 2\right) ,\psi \left( 3\right)
\right) =\left( -\phi \left( 3\right) ,\phi \left( 2\right) ,-\phi \left(
1\right) ,\phi \left( 0\right) \right) $. We can see that the Haar wavelet
obeys the same pattern as the Daubechies wavelet:\ $\psi _{H}=\left( -\phi
\left( 1\right) ,\phi \left( 0\right) \right) $. This pattern can be
extended, using different coefficients but the same general sign rules, for
wavelets with $6$, $8$, $10$, $\ldots $ coefficients. The resulting wavelet
and scaling functions become increasingly smooth, and therefore are better
suited for data sets in which there is only a gradual change in the data set
with position---in thermodynamic systems, this would be more useful for,
say, a spin-$N$ Ising model than a spin-$\frac{1}{2}$ Ising model (presuming
that $N\gg \frac{1}{2}$).

\subsection{Matrix formulation of the wavelet transform}

For discrete systems, a conceptually simple method of implementing the
wavelet transform is to set up the transform as a matrix equation. The input 
$u$ is converted into a column vector ${\bf u},$ so that the coefficients $%
s\left( i\right) $ and $\delta\left( i\right) $ are obtained via the dot
product of ${\bf u}$ with vectors ${\bf h}\left( i\right) $ and ${\bf l}%
\left( i\right) $, where the vectors are padded so that the first nonzero
element is located at position $i$: 
\begin{align*}
{\bf h}\left( i\right) & =\left( 0,\ldots,0,\phi\left( 0\right) ,\phi\left(
1\right) ,\ldots,\phi\left( r-1\right) ,0,\ldots,0\right) , \\
{\bf l}\left( i\right) & =\left( 0,\ldots,0,\psi\left( 0\right) ,\psi\left(
1\right) ,\ldots,\psi\left( r-1\right) ,0,\ldots,0\right) .
\end{align*}
The vectors ${\bf s}$ and ${\bf \delta}$, which contain the
wavelet-transformed coefficients of the decomposition, can be obtained by
forming the matrices ${\bf H}$ and ${\bf L}$ and right-multiplying by the
vector ${\bf u}$: 
\[
{\bf s}={\bf Hu}\text{ and }{\bf \delta}={\bf Lu}\text{,} 
\]
where the rows of the matrices ${\bf H}$ and ${\bf L}$ are the vectors $%
\left( {\bf h}\left( 1\right) ,\ldots,{\bf h}\left( n\right) \right) $ and $%
\left( {\bf l}\left( 1\right) ,\ldots,{\bf l}\left( n\right) \right) $,
respectively. As mentioned in the previous section, we do not need to keep
all of the $s\left( i\right) $'s and $\delta\left( i\right) $'s in order to
obtain perfect reconstruction of our signal; thus, we can obtain all the
necessary coefficients in a single matrix multiplication by combining the
relevant rows of ${\bf H}$ and ${\bf L}$ into a single matrix ${\bf W}%
^{\left( 1\right) }$, whose rows are $\left( {\bf h}\left( 1\right) ,{\bf h}%
\left( 3\right) ,\ldots ,{\bf h}\left( n-1\right) ,{\bf l}\left( 1\right) ,%
{\bf l}\left( 3\right) ,\ldots,{\bf l}\left( n-1\right) \right) $. Thus, the
wavelet transformation (\ref{def}) can be written as 
\begin{equation}
\left[ 
\begin{array}{c}
s\left( 1\right) \\ 
\vdots \\ 
s\left( n-1\right) \\ 
\delta\left( 1\right) \\ 
\vdots \\ 
\delta\left( n-1\right)
\end{array}
\right] =\left[ 
\begin{array}{c}
{\bf h}\left( 1\right) \\ 
{\bf \vdots} \\ 
{\bf h}\left( n-1\right) \\ 
{\bf l}\left( 1\right) \\ 
\vdots \\ 
{\bf l}\left( n-1\right)
\end{array}
\right] \left[ 
\begin{array}{c}
u\left( 1\right) \\ 
u\left( 2\right) \\ 
\\ 
\vdots \\ 
\\ 
u\left( n\right)
\end{array}
\right] .  \label{7}
\end{equation}
We will denote the product on the left-hand side of (\ref{7}) as ${\bf 
\tilde{u}}$.

As stated above, the wavelet process can be applied recursively: the set of
averages $\left( s\left( 1\right) ,s\left( 3\right) ,\ldots ,s\left(
n-1\right) \right) $ can be treated as a new data sample ${\bf u}^{\left(
1\right) }=\left( u^{\left( 1\right) }\left( 1\right) ,\ldots ,u^{\left(
1\right) }\left( n/2\right) \right) $, and operated on by an $\frac{N}{2}%
\times \frac{N}{2}$ reduction of ${\bf W}^{\left( 1\right) }{\bf ,}$ which
we denote ${\bf W}^{\left( 2\right) }$, to produce a new set of $n/4$
averages $\left( s^{\left( 2\right) }\left( 1\right) ,s^{\left( 2\right)
}\left( 3\right) ,\ldots ,s^{\left( 2\right) }\left( n/2-1\right) \right) $
and corresponding new set of $n/4$ differences $\left( \delta ^{\left(
2\right) }\left( 1\right) ,\delta ^{\left( 2\right) }\left( 3\right) ,\ldots
,\allowbreak \delta ^{\left( 2\right) }\left( n/2-1\right) \right) $. To
reconstruct the original data set, we combine these $n/2$ values along with
the $n/2$ differences $\left( \delta ^{\left( 1\right) }\left( 1\right)
,\allowbreak \delta ^{\left( 1\right) }\left( 3\right) ,\ldots ,\allowbreak
\delta ^{\left( 1\right) }\left( n-1\right) \right) $ obtained from applying 
${\bf W}^{\left( 1\right) }$. This process can be repeated as many times as
desired, dividing the $m$ non-downsampled averages ${\bf s}^{\left( k\right)
}$ into $m/2$ averages ${\bf s}^{\left( k+1\right) }$ and $m/2$ differences $%
{\bf \delta }^{\left( k+1\right) }$. However, since at each iteration the
matrix ${\bf W}^{\left( k\right) }$ only operates on selected elements of
the vector ${\bf \tilde{u}}^{\left( k\right) }={\bf W}^{\left( k-1\right) }%
{\bf u}^{\left( k-1\right) }$, computations for multiple levels can be
performed at the same time. Thus, if we wish to apply the wavelet transform $%
K$ times, we can write this operation as an extended matrix product:\cite
{Gines98} 
\begin{equation}
{\bf \tilde{u}}^{\left( K\right) }={\bf Wu}=\prod_{k=1}^{K}{\bf Q}^{\left(
k\right) }{\bf u}\text{,}  \label{10}
\end{equation}
where the ${\bf Q}^{\left( k\right) }$ are a family of matrices of the form 
\begin{equation}
{\bf Q}^{\left( k\right) }=\left[ 
\begin{array}{cc}
{\bf W}^{\left( k\right) } & {\bf 0} \\ 
{\bf 0} & {\bf I}
\end{array}
\right] .  \label{13}
\end{equation}
In (\ref{13}), ${\bf Q}^{\left( k\right) }$ is always an $N\times N$ matrix,
while the matrix ${\bf W}^{\left( k\right) }$ has size $\left(
N/2^{k-1}\right) \times \left( N/2^{k-1}\right) $.

To recover the original data sample ${\bf u}$ following a wavelet transform $%
{\bf W}$, we can simply multiply ${\bf \tilde{u}}$ by the inverse of the
wavelet matrix ${\bf W}$. The matrix ${\bf W}$ is unitary: that is, its
inverse ${\bf W}^{-1}$ is equal to its transpose ${\bf W}^{T}$.
Consequently, if ${\bf W}$ is known, all that is necessary to reverse the
transformation is to left-multiply ${\bf \tilde{u}}^{\left( k\right) }$ by
the transpose ${\bf W}^{T}$. Moreover, the computation (\ref{7}) of the
wavelet transform can usually be carried out ``in place'' by manipulating
local coordinates; in this manner, the computation is carried out even more
rapidly than a standard multiplication, and without the increased storage
costs associated with matrix multiplications.\cite{Golub96,Trefethen97}

\subsection{Multidimensional wavelet transforms}

Since virtually all problems in lattice thermodynamics are in multiple
dimensions, it is necessary to take the wavelet transform of a multivariate
function or data set. Several methods have been developed to carry out such
transformations; among them are Cohen and Daubechies's {\em separable
wavelets}, which form the multidimensional scaling and wavelet functions $%
\phi \left( x,y\right) $, $\psi _{xx}\left( x,y\right) $, $\psi _{xy}\left(
x,y\right) $, and $\psi _{yy}\left( x,y\right) $ from products of the
one-dimensional scaling and wavelet functions $\phi \left( x\right) $ and $%
\psi \left( x\right) $ \cite{Cohen93}. A more general algorithm, the {\em %
lifting algorithm}, has been developed by Sweldens.\cite
{Sweldens97,Daubechies98} It divides the wavelet transform into two steps:\
the first computes the wavelet coefficients $\delta \left( i\right) ;$ the
second step uses the wavelet coefficients to speed up the calculation of the
scaling coefficients. ``Initialization'' of the lifting algorithm requires
the use of an appropriately selected basis function.

A particularly convenient basis function for the multidimensional lifting
transform is the generalized orthogonal Haar wavelets outlined by\ Sweldens. 
\cite{Sweldens97} An extension of the one-dimensional wavelet transform,
they can be created in any number of dimensions, and have the same basic
orthonormality properties as the one-dimensional Haar functions, although
the orthonormality constant becomes $2^{-d/2},$ where $d$ is the
dimensionality of the system. [The two-dimensional version is shown in
Figure \ref{fig2}.] Moreover, the use of the Haar wavelets as a starting
point for further iterations of the lifting algorithm allow the development
of additional, ``better'' lifted wavelets with more desirable properties,
such as smoothness.

It can further be seen that the ``oversampling'' problem which exists in one
dimension will be magnified in multiple dimensions: since the number of
wavelet functions which are produced from each data point increases by a
factor of two with each additional dimension added, we must reduce the
number of points maintained for each wavelet function by that same factor.\
Thus, in two dimensions, we keep only every fourth point; in three
dimensions, every eighth point, and so on. The wavelet transform for
multidimensional systems can thus still be written in the form of (\ref{10})
and (\ref{13}), after we have written the multidimensional data set in terms
of a column vector. This can be accomplished by wrapping around the edges of
the matrix in creating ${\bf u}$: for example, after inserting element $%
\left( 1,N\right) $ of a two-dimensional data set into ${\bf u}$, we next
store element $\left( 2,1\right) $, and so forth. The other significant
difference in the structure of these equations is that the size of the
submatrix ${\bf W}^{\left( k\right) }$ in (\ref{13}) is now $N/2^{\left(
k-1\right) d}\times $ $N/2^{\left( k-1\right) d}$ instead of $%
N/2^{k-1}\times N/2^{k-1}$.

\section{Wavelet analysis of lattice thermodynamics}

\subsection{Applying the wavelet transform to Hamiltonians}

\label{WTLS}The standard model for studying the thermodynamic behavior of
lattice systems is the spin-$\frac{1}{2}$ Ising model, which contains both
nearest-neighbor pairwise interactions as well as interactions between
lattice sites and an external field. The Hamiltonian for this system is
normally written in the form 
\begin{equation}
-\beta {\cal H}=h\sum_{i}\sigma _{i}+J\sum_{\left\langle ij\right\rangle
}\sigma _{i}\sigma _{j},  \label{1}
\end{equation}
where $h$ is the strength of the external field in the direction of the
spins $\sigma _{i}$, and $J$ is the strength of the interaction between
nearest-neighbor pairs of spins on the lattice; these pairs are indicated by
the subscript $\left\langle ij\right\rangle $ in the second summation in (%
\ref{1}). The inverse temperature $\beta =\left( k_{B}T\right) ^{-1}$; for
convenience we let $k_{B}=1$, so that temperature, external field, and
nearest-neighbor interactions are all dimensionless quantities. The model
can be further extended by the inclusion of a position-dependent external
field, or by the inclusion of pairwise interactions beyond nearest
neighbors; in this case, (\ref{1})\ can be written in the more general form 
\begin{equation}
-\beta {\cal H}=\sum_{i}h_{i}\sigma _{i}+\sum_{i}\sum_{j}J_{ij}\sigma
_{i}\sigma _{j},  \label{2}
\end{equation}
where in (\ref{2}) $h_{i}$ is the strength of the external field at lattice
site $i$, and $J_{ij}$ is the strength of the interaction between sites $i$
and $j$. Analytical solutions of (\ref{1}) are well-known for
one-dimensional systems and two-dimensional systems when $h=0$;\cite
{Pathria96,Onsager44} it has recently been suggested that analytic solutions
do not exist for more complicated systems.\cite{Istrail00}

While (\ref{1})\ and (\ref{2}) are compact representations of the
Hamiltonian of the system, the expansion of the lattice variables $s_{i}$
and $s_{j}$ as a sum of wavelet coefficients makes these equations
impractical for applying the wavelet transformation. Since the system is
described discretely, we want to use discrete wavelets, and therefore a
matrix formulation of the Hamiltonian would be convenient. Using graph
theory,\cite{Cormen90} this is readily accomplished:\ let the vectors ${\bf u%
}=\left( \sigma _{1},\sigma _{2},\ldots ,\sigma _{N}\right) $ and ${\bf h}%
=\left( h_{1},\ldots ,h_{N}\right) $ denote the values of each of the $N$
lattice variables in the system and the set of external-field interaction
strengths, respectively, constructed in the row-wise manner described in the
previous section. Furthermore, define the matrix ${\bf J}$ such that element 
$J_{ij}$ is the strength of the interaction between site $i$ and site $j$.
If these sites do not interact, then $J_{ij}=0$. Then, the Hamiltonian (\ref
{2})\ can be written in the form of a matrix equation: 
\begin{equation}
-\beta {\cal H}={\bf h}^{T}{\bf u}+{\bf u}^{T}{\bf Ju},  \label{3}
\end{equation}
where the superscript $T$ denotes the transpose of the vector (or matrix)\
which precedes it.

The matrix ${\bf W}$ which defines the wavelet transform satisfies by
construction ${\bf W}^{T}{\bf W}={\bf I}$, where ${\bf I}$ is the identity
matrix. Therefore, to apply the wavelet transform, we simply insert ${\bf W}%
^{T}{\bf W}$ between each pair of terms in (\ref{3}), thereby obtaining 
\begin{equation}
-\beta {\cal H}=\left( {\bf h}^{T}{\bf W}^{T}\right) \left( {\bf Wu}\right)
+\left( {\bf u}^{T}{\bf W}^{T}\right) \left( {\bf WJW}^{T}\right) \left( 
{\bf Wu}\right) .  \label{4}
\end{equation}
Using the general matrix property that ${\bf B}^{T}{\bf A}^{T}=\left( {\bf AB%
}\right) ^{T}$, (\ref{4}) can be rewritten in terms of the
wavelet-transformed vectors ${\bf \tilde{h}}^{\left( K\right) }{\bf =Wh}$, $%
{\bf \tilde{u}}^{\left( K\right) }{\bf =Wu}$, and the wavelet-transformed
matrix ${\bf \tilde{J}}^{\left( K\right) }{\bf =WJW}^{T}$: 
\begin{equation}
-\beta {\cal \tilde{H}}=\left( {\bf \tilde{h}}^{\left( K\right) }\right) ^{T}%
{\bf \tilde{u}}^{\left( K\right) }+\left( {\bf \tilde{u}}^{\left( K\right)
}\right) ^{T}{\bf \tilde{J}}^{\left( K\right) }{\bf \tilde{u}^{\left(
K\right) }.}  \label{5}
\end{equation}
It is important to note that in writing equation (\ref{5}), we have not made
any explicit assumptions about the form of the matrix ${\bf W}$, other than
to require it to be a matrix describing a wavelet transform. As a result, we
can perform several levels of multiresolution simultaneously through
appropriate preparation of the matrix ${\bf W}$ in the manner outlined
above. Using the inverse wavelet transform ${\bf W}^{-1}={\bf W}^{T}$ to
recover an original configuration ${\bf u}$ is a one-to-one mapping only if
we are provided with $n$ distinct data points, as contained in ${\bf \tilde{u%
}}^{\left( K\right) }$.

Examining (\ref{5}), we see that the result of applying the wavelet
transform to a set of spins ${\bf u}$ is to create a new representation $%
{\bf \tilde{u}}^{\left( K\right) }$, which contains the same information
about the spins as does the original state vector ${\bf u}$. However, the
vector ${\bf \tilde{u}}^{\left( K\right) }$ contains $n/2^{dK}$ averages,
where $d$ is the dimensionality of the lattice; these averages can be viewed
as ``block spins'' in a sense similar to that of Kadanoff.\cite{Goldenfeld92}
The remaining elements of ${\bf \tilde{u}}^{\left( K\right) }$ contain the
local differences in the spins; that is, they can be used to describe the
specific set of spins which give rise to a particular block spin $\tilde{s}%
_{i}^{\left( K\right) }$.

\subsection{Computing thermodynamic functions}

The canonical partition function 
\begin{equation}
Z=\sum_{{\bf u}\in{\Bbb S}}\exp\left( -\beta{\cal H}\left( {\bf u}\right)
\right) ,  \label{6}
\end{equation}
where ${\Bbb S}$ is the configuration space of the system, can be used to
derive all the thermodynamic properties of a lattice system. Applying the
wavelet transform ${\bf W}$ to the state vector ${\bf u}$ results in a new
state vector ${\bf \tilde{u}}^{\left( K\right) }$ belonging to the
configuration space ${\Bbb \tilde{S}}^{\left( K\right) }$. Provided that $%
{\bf W}$ satisfies the perfect reconstruction property, if the summation
over ${\bf u}\in{\Bbb \tilde{S}}$ in (\ref{6}) is replaced with the
summation over ${\bf \tilde{u}}^{\left( K\right) }\in{\Bbb \tilde{S}}%
^{\left( K\right) }$, the results will be identical. This result follows
naturally, since perfect reconstruction necessarily implies that there is a
unique state vector ${\bf \tilde{u}}^{\left( K\right) }\in{\Bbb \tilde {S}}%
^{\left( K\right) }$ for each state vector ${\bf u}\in{\Bbb S}$, and by
construction of the wavelet-transformed Hamiltonian (\ref{5}), $-\beta{\cal H%
}=-\beta{\cal \tilde{H}}^{\left( K\right) }.$

Ensemble averages involving wavelet-transformed variables are in general no
more complicated than computations involving the original variables. In
general, the transformation of a function (or functional) of the original
variables $f\left( {\bf u}\right) $ or $f\left[ u\right] $ will generally
have the same characteristics after applying the wavelet transform to obtain 
$\tilde{f}^{\left( K\right) }\left( {\bf \tilde{u}}^{\left( K\right)
}\right) $ or $\tilde{f}^{\left( K\right) }\left[ \tilde {u}^{\left(
K\right) }\right] $. Moreover, the standard properties of ensemble
averaging, such as linearity, also apply, which makes calculations of
moments of the distribution particularly simple.

The above formulation can be applied to any system whose Hamiltonian can be
written in the form of equation (\ref{2}), or as the sum of contributions,
each of which is of that form. While the wavelet transform can be applied to
any lattice system whose Hamiltonian is a function of the components of the
state vector ${\bf u}$---or, indeed, to any Hamiltonian which is a
functional of ${\bf u}\left( {\bf r}\right) $ for continuous systems---in
most of these cases, it is necessary to rely on the more cumbersome series
expansions, or integral transforms in the case of continuous systems.

In addition, for Ising spin variables, the use of the wavelet transform
poses an additional challenge. While it is entirely straightforward to
describe the possible values an individual lattice site can take---for a
spin-$q$ Ising model, the allowed values of an individual spin $\sigma $ are 
$-q$, $-q+1,\ldots ,q-1,q$---the rules which determine whether an
arbitrarily selected ``transformed'' state vector ${\bf \tilde{u}}^{\left(
K\right) }$ represents a real state vector ${\bf u}$ are cumbersome to
manipulate and to implement, and have proven a formidable challenge in prior
research as well.\cite{Best94}

\section{Analysis}

In this section, we use two variants of the two-dimensional Ising model as a
basis for our calculations: we look at $4\times 4$ and $32\times 32$ Ising
lattices; the former is used for calculations when it is desired to perform
calculations over all states explicitly, while the latter is illustrative of
larger systems, for which exact calculations are intractable. The emphasis
in this paper will be on the use of wavelets to yield approximate answers in
significantly faster time than is possible with a Monte Carlo simulation
incorporating all degrees of freedom explicitly. The methodology by which
the wavelet transform can be extended to Monte Carlo simulations of lattice
systems is the focus of the companion paper.\cite{Ismail02b}

Before proceeding to the results of the calculations, we make note of the
time required to execute the simulations. Each of Figures \ref{fig6} through 
\ref{fig8} plots the observed variables in the temperature range $T=0.50$ to 
$T=5.00$, with a step size of $\Delta T=0.01$. The computations for the
original problem, with $2^{16}=65536$ states to consider explicitly,
required more than 6.8 seconds per point to execute the required
calculations on a 733 MHz Pentium III; the same calculations using one and
two applications of the wavelet transform required just 0.061 and 0.026
seconds per point to consider $5^{4}=625$ and $17$ states, respectively.

\subsection{Weighting functions for wavelet-transformed statistics}

Until now, we have worked with wavelet transforms which preserve the number
of degrees of freedom between the original and transformed problems. This
approach yields results for the wavelet-transformed system in exact
agreement with those for the original system. However, as mentioned above,
in such cases the transformed equations are usually harder to model than the
original ones. Hence it is desirable to use the wavelet transform not only
as a means of describing a lattice system, but also to derive an
approximation scheme whereby estimates of thermodynamic properties can be
made efficiently, while still offering error estimates that bound the true
results.

The approach we adopt in this paper is to ignore all local differences:\
that is, we assume that $\delta_{i}^{\left( k\right) }=0$ for all values of $%
i$ and $k$. As a secondary assumption, we assume that correlation functions
which include wavelet coefficients are also equal to zero: that is, we
assume $\left\langle \delta_{i}^{\left( k\right) }A\left( \cdot\right)
\right\rangle =0$ for any choice of property $A\left( \cdot\right) $. These
extreme assumptions represent a ``worst-case scenario'' for the use of the
wavelet transform method; any more accurate representation of the behavior
of the wavelet coefficients will lead to similarly more accurate results in
the calculations we present below. Note that under certain circumstances,
these approximations are accurate: for Hamiltonians of the form (\ref{2})
which do not exhibit interactions with an external field, there exist
configurations with equal energies but opposite signs for $%
\delta_{i}^{\left( k\right) }$; consequently, when we take the average over
all configurations, the ensemble average of $\delta_{i}^{\left( k\right) }$
will vanish.

Let us assume that we have applied the wavelet transform to our original
Hamiltonian, ${\cal H}\left( {\bf s}\right) ,$ and have obtained a new
function ${\cal \tilde{H}}^{\left( K\right) }\left( {\bf \tilde{u}}^{\left(
K\right) }\right) $. If we then make our approximation that all the wavelet
coefficients are negligible, then we can reduce ${\cal \tilde{H}}^{\left(
K\right) }\left( {\bf \tilde{u}}^{\left( K\right) }\right) $ to a new
function ${\cal \tilde{H}}^{\left( K\right) }\left( {\bf \tilde{s}}^{\left(
K\right) }\right) $, where ${\bf \tilde{s}}^{\left( K\right) }$ represents
the set of all averages $\left( s_{1}^{\left( K\right) },\ldots
,s_{m}^{\left( K\right) }\right) $ that were preserved by the wavelet
transform. We know that the partition function of the system before we
performed the wavelet transform is given by (\ref{6}); after using the
wavelet transform, we hypothesize that the partition function of the new
system is given by 
\begin{equation}
\tilde{Z}=\sum_{{\bf \tilde{s}}^{\left( K\right) }\in {\Bbb \tilde{S}}%
^{\left( K\right) }}w\left( {\bf \tilde{s}}^{\left( K\right) }\right) \exp
\left( -\beta {\cal \tilde{H}}^{\left( K\right) }\left( {\bf \tilde{s}}%
^{\left( K\right) }\right) \right) ,  \label{8}
\end{equation}
where the $w\left( {\bf \tilde{s}}^{\left( K\right) }\right) $ is the weight
of configuration ${\bf \tilde{s}}^{\left( K\right) }$ in the configuration
space ${\Bbb \tilde{S}}^{\left( K\right) }$; since multiple configurations $%
{\bf u}\in {\Bbb S}$ can correspond to the same ${\bf \tilde{s}}^{\left(
K\right) }$, we cannot set $w\left( {\bf \tilde{s}}^{\left( K\right)
}\right) =1$ as we did in the untransformed Ising model. Clearly, the new
partition function (\ref{8}) will be identical to the original partition
function (\ref{6}) if we define 
\begin{equation}
w\left( {\bf \tilde{s}}\right) =\sum_{{\bf u}:{\bf u\rightarrow \tilde{s}}%
^{\left( K\right) }}\exp \left( -\beta \left( {\cal H}\left( {\bf u}\right) -%
{\cal \tilde{H}}^{\left( K\right) }\left( {\bf \tilde{s}}^{\left( K\right)
}\right) \right) \right) ,  \label{9}
\end{equation}
where the notation ${\bf u}:{\bf u}\rightarrow {\bf \tilde{s}}^{\left(
K\right) }$ denotes that the sum is to be performed over all configurations $%
{\bf u}$ which have the same set of wavelet-transformed averages ${\bf 
\tilde{s}}^{\left( K\right) }$. However, evaluating (\ref{9}) to obtain the
weighting functions is no more tractable than the computation of the
original partition function. Thus, any computational efficiency to be gained
is by finding an economical approximation for (\ref{9}). One such approach
is to take $w\left( {\bf \tilde{s}}^{\left( K\right) }\right) $ to be equal
to the number of states ${\bf s}$ whose wavelet transform yields ${\bf 
\tilde{s}}^{\left( K\right) }$, so that (\ref{8}) is the standard form of
the canonical partition function for a system with degenerate energy levels.
Let us call the number of such states ${\bf s}$ with equivalent averages $%
{\bf \tilde{s}}^{\left( K\right) }$ the degeneracy of state ${\bf \tilde{s}}%
^{\left( K\right) }$, and denote this degeneracy as $g\left( {\bf \tilde{s}}%
^{\left( K\right) }\right) $. The restriction of ${\bf \tilde{u}}^{\left(
K\right) }$ to the averages ${\bf \tilde{s}}^{\left( K\right) }$ prevents a
unique reconstruction of ${\bf u},$ unless $g\left( {\bf \tilde{s}}^{\left(
K\right) }\right) =1.$

\subsection{Order parameter}

A natural variable to compute is the order parameter, generically denoted $%
\eta $, which for lattice spin systems is the average magnetization, 
\begin{equation}
m=\frac{1}{N}\left\langle \sum_{i}\sigma _{i}\right\rangle ;  \label{11}
\end{equation}
for other members of the Ising universality class, the order parameter can
represent the overall density $\rho $ or the difference between the
densities of two phases.\cite{Stanley71} While the computation of the order
parameter is straightforward in simulations, its calculation can be made
difficult in the case of zero external field because of symmetries in the
configuration space:\ for every configuration with magnetization $m_{i}$,
there exists another configuration with the same energy and magnetization $%
-m_{i}$. As a result, when all the states are combined using (\ref{11}), we
do not observe spontaneous magnetization, but instead find $m=0$ at all
temperatures. Thus, we consider the absolute value of the magnetization, $%
\left| m\right| $ in place of the magnetization $m$. The results of this
calculation for the $4\times 4$ Ising model with the wavelet transform,
setting all $\delta \left( i\right) $'s to zero, and without the wavelet
transform are shown in\ Figure \ref{fig6}. We note that the error is
essentially negligible for temperatures below $T=1$, and decreases again for
large values of $T$, where differences in energy levels become negligible
and the average magnetization of a state is the only contributing factor to $%
\left| m\right| $.

\subsection{Free-energy considerations}

The Helmholtz free energy is just the logarithm of the partition function: $%
A=-k_{B}T\ln Z$; if we estimate the partition function using an expression
such as (\ref{8}), we naturally expect the approximated value $A$ to differ
from its true value. When we examine the behavior of the $4\times 4$ Ising
model under the wavelet transform, we find as expected that the two free
energy surfaces are similar, although they are clearly not identical. In
particular, the exact numerical values obtained from the two equations are
not the same. However, since the assignment $A=0$ is arbitrary in any
system, we can choose to define $A=0$ as either the maximum or the minimum
free energy obtained in each system. Under these conditions, the energy
scales for the exact calculation and the wavelet transform calculation are
essentially identical, particularly at the so-called ``fixed points'' of the
system---that is, for points which are not affected by renormalization
transformations.\cite{Ma76} For the two-dimensional Ising model, these
points are at $T=0$, and at infinite (positive or negative) values of the
external field interaction $h$.

This agreement at the fixed points should not be surprising: the fixed
points correspond to unique configurations of the system, such that for the
configurations ${\bf \tilde{s}}^{\left( K\right) }$ which result from the
wavelet transform of the fixed points ${\bf u}^{\ast}$, the degeneracy of
the states is unity. Consequently, the approximation $w\left( {\bf \tilde {s}%
}^{\left( K\right) }\right) =g\left( {\bf \tilde{s}}^{\left( K\right)
}\right) $ is correct for the dominant configuration in the system, and
therefore the behavior of the approximate partition function $Z^{\prime}$ is
almost identical to that of the true partition function $Z$ in the vicinity
of the fixed points.

As can be seen in Figure \ref{fig7}, the free energy converges to the same
values in the low-temperature limit, where only a few states\ which are
essentially unaffected by the wavelet transform contribute to the partition
function of the system. At high temperatures, the agreement is less exact,
because of the approximations made in evaluating the Boltzmann factors of
the block spin configurations.

\subsection{Entropy of a wavelet-transformed lattice system}

It is relatively straightforward to show that coarse-graining a system with
a countable phase space---that is, a phase space whose states can be
completely enumerated---requires a correction factor to ensure satisfactory
agreement with results from the fine-scale system. As an example, consider
the entropy of a wavelet-transformed spin-$\frac{1}{2}$ Ising model. The
original lattice has $2^{N_{t}}$ total configurations, and therefore in the
high-temperature limit, the entropy approaches 
\begin{equation}
S_{\max }=k_{B}N_{t}\ln 2.  \label{e6}
\end{equation}
If we ignore the weighting factor of the wavelet-transformed system by
setting $w\left( {\bf \tilde{s}}^{\left( K\right) }\right) =1$ for all
states ${\bf \tilde{s}}^{\left( K\right) }$, the resulting system will have $%
\left( 1+\prod_{i=1}^{K-1}N^{\left( i\right) }\right) ^{N^{\left( K\right)
}} $, and the maximum possible entropy for this system is 
\begin{equation}
S_{\max }^{\left( K\right) }=k_{B}N^{\left( K\right) }\ln \left(
\prod_{i=1}^{K-1}N^{\left( k\right) }+1\right) .  \label{e1}
\end{equation}
Comparing (\ref{e1}) to the original limit of $S_{\max }=k_{B}N_{t}\ln 2$,
we must have $S_{\max }>S_{\max }^{\left( K\right) }$, since the
configuration space is smaller, which indicates that our coarse-graining has
effectively reduced the available entropy of the system. We can conclude
from this that unless the volume of phase space is conserved by the
coarse-graining procedure, some entropy will be lost at high temperatures,
regardless of the accuracy of the coarse-graining. Our choice of the
degeneracy $g\left( {\bf \tilde{s}}^{\left( K\right) }\right) $ for the
weighting function $w\left( {\bf \tilde{s}}^{\left( K\right) }\right) $
preserves the volume of phase space in the limit of $T\rightarrow \infty $
for the systems under study here.

Looking at the specific case of the $4\times 4$ Ising model, we compute the
entropy as a function of temperature for zero external field for the exact
problem, and for the wavelet-transformed problem using one and two
iterations of the two-dimensional Haar wavelet. The results are shown in
Figure \ref{fig5}. As before, we find that errors vanish in the
low-temperature limit, where only the lowest-energy states make a
contribution to the partition function; since there are two such states,
namely those with all spins up and all spins down, we find that the $%
T\rightarrow 0$ limit of the entropy is 
\[
S\left( T\rightarrow 0\right) \approx -\sum_{i=1}^{2}\frac{1}{2}\ln \frac{1}{%
2}=\ln 2 
\]
In the high-temperature limit, the entropies agree because, by construction,
we have $\sum_{{\bf \tilde{s}\in }{\Bbb \tilde{S}}}w\left( {\bf \tilde{s}}%
\right) =2^{16},$and therefore, since the Boltzmann factor goes to unity for
all states as $T\rightarrow \infty $, $Z\rightarrow \sum_{{\bf \tilde{s}\in }%
{\Bbb \tilde{S}}}w\left( {\bf \tilde{s}}\right) $, and the entropy tends
toward the value $S_{\max }$ given by (\ref{e6}). Disagreement in the
intermediate temperature regime is largely the result of grouping together
states with different energies into a single transformed state with a single
energy. In particular, the grouping of higher-energy states together at a
lower energy increases the probability of being in those states, and
therefore increases the total entropy of the system. This explains the
seemingly anomalous result of coarse-graining increasing the entropy at
intermediate temperatures observed in Figure \ref{fig5}.

For more general systems, we can consider the formal definition of the
entropy for a continuous distribution, 
\begin{equation}
S=-k_{B}\int_{{\bf u}\in {\Bbb S}}d{\bf u}\,p\left( {\bf u}\right) \ln
p\left( {\bf u}\right) ,  \label{e2}
\end{equation}
where $p\left( {\bf u}\right) $ is the probability of observing the
configuration ${\bf u}$. For the original Ising model, each configuration $%
{\bf u}$ is unique, and therefore the probability $p\left( {\bf u}\right) $
is just the Boltzmann weight $Z^{-1}\exp \left( -\beta {\cal H}\left( {\bf u}%
\right) \right) ,$ and thus the entropy is given by 
\begin{equation}
S=k_{B}\ln Z+\frac{k_{B}}{Z}\int_{{\bf u}\in {\Bbb S}}d{\bf u}\beta {\cal H}%
\left( {\bf u}\right) \exp \left( -\beta {\cal H}\left( {\bf u}\right)
\right) .  \label{e3}
\end{equation}
After performing the wavelet transform, and discarding finer-scale details,
we must account for the weighting factor $w\left( {\bf \tilde{s}}\right) $
in our expression for the entropy. Consequently, the entropy for such a
system is given by 
\begin{eqnarray}
S^{\left( K\right) } &=&-k_{B}\int_{{\bf \tilde{s}\in }{\Bbb \tilde{S}}}d%
{\bf \tilde{s}\,}w\left( {\bf \tilde{s}}\right) p\left( {\bf \tilde{s}}%
\right) \ln p\left( {\bf \tilde{s}}\right)  \nonumber \\
&=&k_{B}\ln \tilde{Z}+\frac{k_{B}}{\tilde{Z}}\int_{{\bf \tilde{s}\in }{\Bbb 
\tilde{S}}}d{\bf \tilde{s}}\,w\left( {\bf \tilde{s}}\right) \beta {\cal 
\tilde{H}}^{\left( K\right) }\left( {\bf \tilde{s}}^{\left( K\right)
}\right) e^{-\beta {\cal \tilde{H}}^{\left( K\right) }\left( {\bf \tilde{s}}%
^{\left( K\right) }\right) },  \label{e4}
\end{eqnarray}
where the coarse-grained partition function $\tilde{Z}$ is given by (\ref{8}%
).

\subsection{Fluctuation properties}

For a given thermodynamic system, the constant-volume heat capacity $C$ is
defined as the fluctuation in the internal energy of the system: 
\begin{equation}
C=\frac{\left\langle E^{2}\right\rangle -\left\langle E\right\rangle ^{2}}{%
k_{B}T^{2}}.  \label{fp1}
\end{equation}
It is well-known that in the vicinity of a continuous phase transition, the
heat capacity $C$ diverges. At the same time, it is also known that no
system of finite size can have $C\rightarrow \infty $, since the energy, and
hence the variance $\left\langle E^{2}\right\rangle -\left\langle
E\right\rangle ^{2}$ in the energy of the system will also remain finite.
However, we can still observe evidence of power-law divergence near the
critical temperature.\cite{Binney93} For small finite systems, we do not
observe evidence of a divergence in the heat capacity; instead, the heat
capacity is a smooth function of temperature $T$. The wavelet transform
largely preserves the behavior of the original model; we observe the same
general functional form in both the original model as well as the
transformed systems, as seen in Figure \ref{fig8}. However, for large finite
systems, we can still detect the characteristic power-law divergence in the
vicinity of the critical point; such a divergence is shown for a $32\times
32 $ Ising model as Figure \ref{fig3}.

Applying the wavelet transform to a given thermodynamic system, we expect
that we will still find evidence of a critical point; however, because of
the mean-field like behavior of the wavelet transform method, the critical
point should be found at a higher temperature in the wavelet-transformed
system than in the original system. Computing the heat capacity for large
lattices cannot be done by exhaustive enumeration; therefore, we save
discussion of our numerical results for a subsequent paper,\cite{Ismail02b}
and present a brief heuristic argument. As the size of a block spin
increases, corresponding to a greater number of iterations of the wavelet
transform, we expect that the suppression of fluctuations in the system will
lead to an increase in the temperature of the phase transition, while the
actual numerical value of the heat capacity itself decreases. The decrease
in the number of degrees of freedom in the system is accompanied by an
increasing trend towards homogeneity of the system---the remaining
configurations begin to look more and more similar to one another. As a
result, the variance measured by $\left\langle E^{2}\right\rangle
-\left\langle E\right\rangle ^{2}$ decreases with increasing amounts of
coarse-graining. This trend is much less pronounced for small systems, as
shown in Figure \ref{fig8}. When the number of degrees of freedom is small,
the ability to perform an exact enumeration ensures that the only errors
observed are those introduced by the wavelet approximation itself. Thus,
since the free energy and entropy of small systems are accurately reproduced
by the transform, only minor deviations in the phase transition temperature
for small systems are expected.

\section{Similarities to Renormalization Group theories}

It should already be apparent that close affinities exist between the method
described here and position-space renormalization methods. The construction
of wavelet-based averages naturally corresponds to Kadanoff's concept of a
``block spin'' transformation:\cite{Goldenfeld92}\ both the present method
and Kadanoff's approach rely on combining a region of contiguous spins into
a single new ``block spin.'' However, with the approach outlined above, we
do not seek to impose the requirement that the block spins must be
restricted to have the same spin values as the original spins. For example,
in a wavelet-transformed spin-$\frac{1}{2}$ Ising model, the block spins can
take on values other than $+1$ and $-1$. In addition, we do not impose
``majority rules''\ or other ``tiebreakers'' in the case where there are
equal numbers of ``up'' and ``down'' spins contained in a single block.

A less apparent connection can also be established with Migdal's
``bond-moving'' approximation,\cite{Migdal76,Migdal76b} or similarly with
Wilson's recursion method.\cite{Wilson74} In particular, Migdal's method can
be compared to applying a separable wavelet transformation, in that the
recursion is performed in only one spatial direction at a time. However,
using the methods of either Migdal or Wilson, after performing the
bond-moving or decimation transformation, the new interaction strengths $%
J_{ij}^{\prime }$ are determined by manipulating the resulting Hamiltonian
and recasting it in the form of the original, leading at times to very
complicated, nonlinear formulas for the $J_{ij}^{\prime }$ as a function of $%
J_{ij}$. By contrast, using wavelets, the values for the transformed
coupling coefficients $\tilde{J}_{ij}$ are obtained directly by transforming
the interaction matrix ${\bf J}$, since ${\bf \tilde{J}}={\bf WJW}^{T}$. The
trade-off that must be made for the algorithmic transparency of obtaining
the new coupling constants via a matrix multiplication is that we cannot
solve for the fixed points of the recursion relation, which makes the
production of a renormalization ``flow diagram'' using the wavelet method a
difficult problem.

The behavior of the wavelet transform and of the renormalization group
differs in another, important manner. By the nature of its construction, the
group of renormalization transformations is only a {\em semi-group},
inasmuch as reversing the mapping to move from a coarsened description to a
more detailed description is impossible.\cite{Gennes79,Fisher98} Using the
wavelet transform, a reverse mapping is theoretically possible:
reversibility fails because of the approximations invoked to reduce the
number of wavelet coefficients which are kept, rather than as an inherent
limitation of the wavelet transform itself.

\section{Conclusions}

Using the wavelet transform as a basis for calculations of lattice systems
yields an impressive reduction in the computational time required to obtain
estimates of thermodynamic properties, at the price of modest errors in
accuracy, given the relatively small size of the systems considered here,
and the approximations introduced to simplify our calculations. The wavelet
transform provides a systematic approach to coarse-grain systems to any
level of resolution. The method produces correct results in the vicinity of
fixed attractors, with decreasing accuracy as one approaches the critical
point in parameter space.

In addition, the use of the wavelet transform on such systems is simple to
implement:\ after selecting a particular set of wavelets, all of the
operations can be reduced to matrix multiplications, as shown in Section \ref
{WTLS}. The computational requirements needed to implement the
transformation are small: if the transformation cannot be accomplished in
the explicit form of a matrix multiplication, it is possible to organize the
calculation to be performed in-place.\cite{Sweldens97} Moreover, the amount
of coarse-graining that can be achieved using the wavelet transform can be
adjusted dynamically. These properties of the wavelet transform lead
naturally to an extension of the method to systems for which simulations are
required; the resulting algorithm is the focus of the companion paper.\cite
{Ismail02b}

\section{Acknowledgments}

Funding for this research was provided in part by a Computational Sciences
Graduate Fellowship (AEI) sponsored by the Krell Institute and the
Department of Energy.


\begin{references}
\bibitem{Freed87}  {\sc K.~F. Freed}, 
\newblock {\em Renormalization Group
Theory of Macromolecules}, \newblock Wiley, New York, 1987.

\bibitem{Rapold96}  {\sc R.~F. Rapold} and {\sc W.~L. Mattice}, 
\newblock
{\em Macromolecules} {\bf 29}, 2457 (1996).

\bibitem{Cho97}  {\sc J.~Cho} and {\sc W.~L. Mattice}, 
\newblock {\em
Macromolecules} {\bf 30}, 637 (1997).

\bibitem{Doruker97}  {\sc P.~Doruker} and {\sc W.~L. Mattice}, 
\newblock
{\em Macromolecules} {\bf 30}, 5520 (1997).

\bibitem{Doruker99}  {\sc P.~Doruker} and {\sc W.~L. Mattice}, 
\newblock
{\em Macromol. Theory Simul.} {\bf 8}, 463 (1999).

\bibitem{Arias99}  {\sc T.~A. Arias}, 
\newblock {\em Rev. Mod. Phys.} {\bf
71}, 267 (1999).

\bibitem{Beylkin99}  {\sc G.~Beylkin}, {\sc N.~Coult}, and {\sc M.~J.
Mohlenkamp}, \newblock {\em J. Comput. Phys.} {\bf 152}, 32 (1999).

\bibitem{Johnson99}  {\sc B.~R. Johnson}, {\sc J.~P. Modisette}, {\sc P.~J.
Nordlander}, and {\sc \ J.~L. Kinsey}, 
\newblock {\em J. Chem. Phys.} {\bf
110}, 8309 (1999).

\bibitem{Luettgen93}  {\sc M.~R. Luettgen}, {\sc W.~C. Karl}, {\sc A.~S.
Willsky}, and {\sc R.~R. Tenney}, 
\newblock {\em IEEE Trans. Sign.
Process.} {\bf 41}, 3377 (1993).

\bibitem{Luettgen93b}  {\sc M.~R. Luettgen}, 
\newblock {\em Image Processing
with Multiscale Stochastic Models}, \newblock PhD thesis, Massachusetts
Institute of Technology, 1993.

\bibitem{Huang97}  {\sc D.-W. Huang}, \newblock {\em Phys. Rev. D} {\bf 56},
3961 (1997).

\bibitem{Gamero97}  {\sc L.~G. Gamero}, {\sc A.~Plastino}, and {\sc M.~E.
Torres}, \newblock {\em Physica A} {\bf 246}, 487 (1997).

\bibitem{Ocarroll93}  {\sc M.~O'Carroll}, 
\newblock {\em J. Stat. Phys.}
{\bf 73}, 945 (1993).

\bibitem{Ocarroll93b}  {\sc M.~O'Carroll}, 
\newblock {\em J. Stat. Phys.}
{\bf 71}, 415 (1993).

\bibitem{Battle99}  {\sc G.~Battle}, 
\newblock {\em Wavelets and
Renormalization}, \newblock World Scientific, Singapore, 1999.

\bibitem{Haar10}  {\sc A.~Haar}, \newblock {\em Math. Ann.} {\bf 69}, 331
(1910).

\bibitem{Mallat89}  {\sc S.~G. Mallat}, 
\newblock {\em IEEE Trans. Pattern
Analysis and Machine Intelligence} {\bf 11}, 674 (1989).

\bibitem{Mallat89b}  {\sc S.~G. Mallat}, 
\newblock {\em Trans. Amer. Math.
Soc.} {\bf 315}, 69 (1989).

\bibitem{Daubechies92}  {\sc I.~Daubechies}, 
\newblock {\em Ten Lectures on
Wavelets}, volume~61 of {\em {CBMS}-{NSF} Regional Conference Series in
Applied Mathematics}, \newblock SIAM, Philadelphia, 1992.

\bibitem{Cohen93}  {\sc A.~Cohen} and {\sc I.~Daubechies}, 
\newblock {\em
Rev. Mat. Ibero-amer.} {\bf 9}, 51 (1993).

\bibitem{Sweldens94}  {\sc W.~Sweldens} and {\sc R.~Piessens}, 
\newblock
{\em SIAM J. Numer. Anal.} {\bf 31}, 1240 (1994).

\bibitem{Sweldens94b}  {\sc W.~Sweldens} and {\sc R.~Piessens}, 
\newblock
{\em Numer. Math.} {\bf 68}, 377 (1994).

\bibitem{Strang96}  {\sc G.~Strang} and {\sc T.~Nguyen}, 
\newblock {\em
Wavelets and Filter Banks}, \newblock Wellesley-Cambridge, Cambridge, MA,
1996.

\bibitem{Sweldens96}  {\sc W.~Sweldens} and {\sc P.~Schr{\"o}der}, \newblock %
Building your own wavelets at home, \newblock in {\em Wavelets in Computer
Graphics}, pp. 15--87, ACM SIGGRAPH Course notes, New Orleans, 1996.

\bibitem{Daubechies88}  {\sc I.~Daubechies}, 
\newblock {\em Commun. Pure
Appl. Math.} {\bf 41}, 909 (1988).

\bibitem{Gines98}  {\sc D.~Gines}, {\sc G.~Beylkin}, and {\sc J.~Dunn}, %
\newblock {\em Appl. Comput. Harmon. A.} {\bf 5}, 156 (1998).

\bibitem{Golub96}  {\sc G.~H. Golub} and {\sc C.~F. Van~Loan}, 
\newblock
{\em Matrix Computations}, \newblock Johns Hopkins University Press,
Baltimore, 1996.

\bibitem{Trefethen97}  {\sc L.~N. Trefethen} and {\sc D.~Bau~III}, 
\newblock
{\em Numerical Linear Algebra}, \newblock SIAM, Philadelphia, 1997.

\bibitem{Sweldens97}  {\sc W.~Sweldens}, 
\newblock {\em SIAM J. Math. Anal.}
{\bf 29}, 511 (1997).

\bibitem{Daubechies98}  {\sc I.~Daubechies} and {\sc W.~Sweldens}, 
\newblock
{\em J. Fourier Anal. Appl.} {\bf 4}, 247 (1998).

\bibitem{Pathria96}  {\sc R.~K. Pathria}, 
\newblock {\em Statistical
Mechanics}, \newblock Butterworth-Heinemann, Woburn, MA, 1996.

\bibitem{Onsager44}  {\sc L.~Onsager}, \newblock {\em Phys. Rev.} {\bf 65},
117 (1944).

\bibitem{Istrail00}  {\sc S.~Istrail}, \newblock Statistical Mechanics,
Three-Dimensionality, and {NP}-Completeness: I. {U}niversality of
Intractability for the Partition Functionof the {I}sing Model Across
Non-Planar Surfaces, \newblock in {\em Proceedings of the 32nd Annual {ACM} {%
S}ymposium on {T}heory of {C}omputing}, pp. 87--96, Portland, OR, 2000, ACM
Press.

\bibitem{Cormen90}  {\sc T.~H. Cormen}, {\sc C.~E. Leiserson}, and {\sc %
R.~L. Rivest}, \newblock {\em Introduction to Algorithms}, \newblock McGraw
Hill-MIT Press, Cambridge, MA, 1990.

\bibitem{Goldenfeld92}  {\sc N.~Goldenfeld}, 
\newblock {\em Lectures on
Phase Transitions and the Renormalization Group.}, \newblock Addison-Wesley,
Reading, MA, 1992.

\bibitem{Best94}  {\sc C.~Best}, {\sc A.~Sch{\"a}fer}, and {\sc W.~Greiner}, %
\newblock {\em Nucl. Phys. B: Proc. Suppl.} {\bf 34}, 780 (1994).

\bibitem{Ismail02b}  {\sc A.~E. Ismail}, {\sc G.~Stephanopoulos}, and {\sc %
G.~C. Rutledge}, \newblock {\em Submitted to J. Chem. Phys.} (2002).

\bibitem{Stanley71}  {\sc H.~E. Stanley}, 
\newblock {\em Introduction to
Phase Transitions and Critical Phenomena}, \newblock Clarendon Press-Oxford,
Oxford, 1971.

\bibitem{Ma76}  {\sc S.-K. Ma}, 
\newblock {\em Modern Theory of Critical
Phenomena}, \newblock Benjamin/Cummings, Reading, MA, 1976.

\bibitem{Binney93}  {\sc J.~J. Binney}, {\sc N.~J. Dowrick}, {\sc A.~J.
Fisher}, and {\sc M.~E.~J. Newman}, 
\newblock {\em The Theory of Critical Phenomena: An Introduction to the
  Renormalization Group}, \newblock Oxford University Press, Oxford, 1993.

\bibitem{Migdal76}  {\sc A.~A. Migdal}, 
\newblock {\em Sov. Phys.-JETP} {\bf
42}, 413 (1976).

\bibitem{Migdal76b}  {\sc A.~A. Migdal}, 
\newblock {\em Sov. Phys.-JETP}
{\bf 42}, 743 (1976).

\bibitem{Wilson74}  {\sc K.~J. Wilson} and {\sc J.~Kogut}, 
\newblock {\em
Phys. Rep.} {\bf 12}, 75 (1974).

\bibitem{Gennes79}  {\sc P.-G. de~Gennes}, 
\newblock {\em Scaling Concepts
in Polymer Physics}, \newblock Cornell University Press, New York, 1979.

\bibitem{Fisher98}  {\sc M.~E. Fisher}, 
\newblock {\em Rev. Mod. Phys.} {\bf
70}, 653 (1998).
\end{references}

\newpage

\begin{figure}[tbh]
\caption{(a) A sample signal $u\left( n\right) $. (b, c) The one-dimensional
Haar scaling function $\protect\phi \left( x\right) $ and wavelet
(differencing) function $\protect\psi \left( x\right) $. (d, e) The
first-level scaling coefficients $s\left( n\right) $ and $\protect\delta %
\left( n\right) $ produced from the original signal $u\left( n\right) $
using the Haar pair, following downsampling of the signal.}
\label{fig1}
\end{figure}

\begin{figure}[tbph]
\caption{The two-dimensional orthogonal Haar wavelets. The coefficient in
all cases is $1/2$, times the sign indicated in each quadrant. Note that the
scaling function $\protect\phi $ is built from the one-dimensional Haar
scaling function, while the wavelet functions are built from the
one-dimensional wavelet function.}
\label{fig2}
\end{figure}

\begin{figure}[tbph]
\caption{(a) Average absolute magnetization $\left| m\right| =\sum_{i}\left| 
\protect\sigma _{i}\right| $ of the $4\times 4$ Ising model at zero external
field as a function of temperature, as computed by an exact enumeration
using no wavelet transform (solid line), and using one and two iterations of
the two-dimensional Haar wavelet (dot-dashed and dotted lines,
respectively). (b) Error in the average absolute magnetization of the $%
4\times 4$ Ising model for one and two iterations of the two-dimensional
Haar wavelet versus an exact enumeration.}
\label{fig6}
\end{figure}

\begin{figure}[tbph]
\caption{Free energy $A$ of the $4\times 4$ Ising model at zero external
field as a function of temperature, as computed by an exact enumeration
using no wavelet transform (solid line), and using one and two iterations of
the two-dimensional Haar wavelet (dashed and dot-dashed lines,
respectively). }
\label{fig7}
\end{figure}

\begin{figure}[tbp]
\caption{Entropy of the $4\times 4$ Ising model at zero external field as a
function of temperature, as computed by an exact enumeration using no
wavelet transform (solid line), and using one and two iterations of the
two-dimensional Haar wavelet (dashed and dot-dashed lines, respectively).
The bottom of the $y$-axis corresponds to the zero-temperature limit of $%
S=\ln 2$.}
\label{fig5}
\end{figure}

\begin{figure}[tbp]
\caption{Heat capacity of the $4\times 4$ Ising model at zero external field
as a function of temperature, as computed by an exact enumeration using no
wavelet transform (solid line), and using one and two iterations of the
two-dimensional Haar wavelet (dashed and dot-dashed lines, respectively). }
\label{fig8}
\end{figure}

\begin{figure}[tbph]
\caption{Heat capacity of a $32\times 32$ Ising model, illustrating the
power-law divergence in the vicinity of the critical point (identified by
the arrow).}
\label{fig3}
\end{figure}

\end{document}